\documentclass[a4paper,fleqn,usenatbib]{mnras}

\usepackage{mathptmx}

\usepackage[T1]{fontenc}
\usepackage{ae,aecompl}

\usepackage{graphicx}	
\usepackage{amsmath}	
\usepackage{amssymb}	

\newcommand{\WID}[2]{#1_{\mathrm{#2}}}
\newcommand{\isn}[2]{\mbox{$^{#2}${#1}}}

\title[Asymmetric Nuclear Light Clusters In Supernova Matter]{Asymmetric Nuclear Light Clusters In Supernova Matter}

\author[A. V. Yudin et al.]{
A. V. Yudin,$^{1}$\thanks{E-mail: yudin@itep.ru}
M. Hempel,$^{2}$
S. I. Blinnikov,$^{1}$
D. K. Nadyozhin,$^{1,3}$
I. V. Panov$^{1,3}$
\\
$^{1}$Institute for Theoretical and Experimental Physics, B. Cheremushkinskaya, 25, 117218 Moscow, Russia\\
$^{2}$Department of Physics, University of Basel, Klingelbergstrasse 82, 4056 Basel, Switzerland\\
$^{3}$National Research Center ``Kurchatov Institute'', Akademika Kurchatova pl. 1, 123182, Moscow, Russia\\
}

\date{Accepted XXX. Received YYY; in original form ZZZ}

\pubyear{2018}

\begin{document}
\label{firstpage}
\pagerange{\pageref{firstpage}--\pageref{lastpage}}
\maketitle

\begin{abstract}
We explore the appearance of light clusters at high densities of collapsing stellar cores.
Special attention is paid to the unstable isotope \isn{H}{4}, which was not included in previous studies.
The importance of light clusters in the calculation of rates for neutrino matter interaction is discussed. The main conclusion is that thermodynamic quantities are only weakly sensitive to the chemical composition. The change in pressure and hence the direct change in collapse dynamics will be minor. But the change in neutrino heating and neutronization processes can be significant.
\end{abstract}




\section{Introduction}

The equation of state (EoS) was one of the major ingredients of supernova modelling since the very beginning.
The first attempts to calculate the process of stellar gravitational collapse used very simplified
EoSs: the matter was considered as an ideal gas of free neutrons, protons, alpha--particles and \isn{Fe}{56}
under the condition of Nuclear Statistical Equilibrium (NSE),
surrounded by electron--positron pairs and blackbody radiation \citep{Imsh&Nad1965}. Later this
approach was modified (see, for example \citet{Langer1969}): instead of using one ``heavy'' nucleus with
fixed parameters, an ``average'' nucleus was considered, representing the full chemical composition. Its parameters
(mass number $A$, charge $Z$ etc.) were determined by minimizing the total free energy of the system. This approach is very
popular, and also applied in the most widely used current EoSs of \citet{Latt&Swesty1991} and \citet{Shen1998a}, \citet{Shen1998b}.
Furthermore, in both of these two EoSs, of all possible light nuclei, only alpha particles are considered.
One of the most intriguing
predictions of this type of EoS is the leading role of very massive nuclei (with mass number $A\gtrsim 100$) at high
densities up to nuclear density $\rho\sim 10^{14}~\mbox{g}\cdot\mbox{cm}^{-3}$. In parallel to this, another approach based on the
concept of NSE was developed \citep{Cliff&Taylor1965}.
It explicitly includes an ensemble of different nuclei (hundreds or even thousands). The properties of nuclei (binding energies,
spins of ground state etc.) are chosen from the experiment or from some model considerations. The advantage of such
an approach is the possibility to correctly describe a rich chemical composition of matter in the regions where
nuclei with well--known properties dominate. On the other hand, the predictions of this type of EoS in high--density
neutronized regions of collapsing stellar cores are not reliable, if the selection of nuclei is too restricted.
Additional uncertainty comes from the modelling of medium modifications of the nuclei, which are more difficult to describe and to implement for an ensemble of nuclei.
One prominent difference of the two types of EoSs
concerns the high--density and finite temperature region: NSE type EoSs predict here a large abundance of light neutron--rich clusters.

There are many EoS models in the literature, which go beyond
the simplification of EoSs of \citet{Shen1998a}, \citet{Shen1998b} and \citet{Latt&Swesty1991} to consider all possible light nuclei only made of alpha particles. For an overview, see also the review about supernova EoSs by \cite{Oertel2016}.
At very low densities, the finite-temperature EoS is given model
independently by the virial EoS.
\cite{Horowitz&Schwenk2006} considered first only neutrons, protons and $\alpha$-particles
as basic constituents and experimental information
on binding energies and phase shifts was used in the calculation of
the second virial coefficients. \cite{Oconnor2007} added
\isn{H}{3} and \isn{He}{3} nuclei, and even heavier species were
included by \cite{Mallik2008}. The most advanced approach for light nuclei and their medium
modifications is given by the quantum statistical model, see \cite{Roepke2015} and references therein. In contrast to the virial EoS
this model is able to describe the dissolution of clusters at high densities.
Some other, more phenomenological approaches will be discussed below.
A general finding, present in all
models that consider an ensemble of different light nuclei, is that deuterons,
tritons and helions appear abundantly for
typical conditions of supernova matter in addition to
$\alpha$-particles, see, e.g.,
\cite{Sumiyoshi&Roepke2008,Typel2010,Hempel&Schaffner2010,Pais2015}.
The presence of light clusters can modify weak interaction rates and therefore
the dynamics of astrophysical processes, see, e.g., the works of \cite{Furusawa2013,Fischer2016}.

Light nuclei are also produced in Fermi-energy heavy-ion collisions.
The measurement of their abundances allows to determine
the density and temperature of warm dilute matter
from experiments, see, e.g., \cite{Kowalski2007,Natowitz2010,Wada2012}.
The thermodynamic conditions are similar
to those in the neutrinosphere of core-collapse supernovae
\cite{Horowitz2014}. This allows to confront predictions of supernova EoS models
with experimentally measured yields, see, e.g., \cite{Hempel2015}, whereas
one has to take into account the systematic differences between matter in
heavy-ion collisions and core-collapse supernovae. Unfortunately, \isn{H}{4}
was not discussed in the aforementioned studies.

This article is devoted to the light clusters aspect of the
high-density EoS problem and especially to the role of the ``forgotten'' isotope \isn{H}{4}.
The paper is organised as follows. In section \ref{sec_EOSes} we discuss different types of EoSs used. Section \ref{sec_4hnse} is devoted to the nuclear properties of isotope \isn{H}{4} and its incorporation
into our calculations. Section \ref{sec_Infall} presents the results
of our calculations for the infall stage of collapse (1~ms before bounce).
The role of nuclear partition functions is discussed in section \ref{sec_varpar}.
Section \ref{sec_varEOSs} deals with the comparison of results for various EoSs.
Section \ref{sec_post_bounce} presents post-bounce calculations (about 200~ms after bounce).
Section \ref{sec_Importance} summarizes the major effects produced by the nuclear light clusters,
and especially the isotope \isn{H}{4}. Concluding remarks are given in section \ref{sec_Conclusions}.

\section{EoS models}
\label{sec_EOSes}
\subsection{Pure NSE EoS}
\label{sec_NSE}
As a base case for our research we use the extended NSE EoS model from \citet{Yud&Nad2004}.
All nuclei are assumed to form ideal gases except free neutrons and protons for which degeneracy effects are also
included. Non--ideal corrections (e.g. Coulomb interaction) are not included. Nuclei parameters (binding energies, known
excited states etc) are taken from the online database of the National Nuclear Data Center, see the connected discussion below in section
\ref{sec_varpar} about nuclei partition function handling. The original set of nuclides from \citet{Yud&Nad2004} is extended
mostly by inclusion of neutron--rich isotopes from hydrogen to iron-peak nuclei. Now the total number of nuclei taken into account is 398.
Beside nuclei, this model accounts for electron--positron pairs of arbitrary degeneracy and relativism and blackbody radiation according to \citet{Blinnikov1996}.

\subsection{HS EoS}
\label{sec_HS}
In the model of the HS EoS \citep{Hempel&Schaffner2010},
nuclei are treated as classical Maxwell-Boltzmann
particles and nucleons as interacting Fermi-Dirac particles
employing different relativistic mean-field parameterizations.
Here we use the version HS(DD2) \citep{Fischer2014} with the parameterization DD2 of \cite{Typel2010}.
This EOS is available in tabular form and covers the full range in density, temperature, and electron fraction,
so that it can be applied in simulations of core-collapse supernovae or neutron star mergers.
Several thousands of nuclei are considered, including light
ones. In the version we are using here, their binding energies are either taken from experimental
measurements \citep{Audi&Wapstra1995} if available or otherwise from the theoretical nuclear
structure calculation of \cite{Moeller1995}.

The following medium modifications are incorporated for nuclei: screening of the Coulomb
energies by the surrounding gas of electrons in Wigner-Seitz approximation
and excluded-volume effects. In addition, excited states are taken into account by
an internal partition function using the level density of \cite{Fai&Randrup1982}.
Since the slightly updated version of the HS EOS was published by \cite{Hempel2012},
the total binding energy of each nucleus is introduced as a cut-off for
its highest possible excitation energy. For the groundstate spin, the following naive
prescription is used: nuclei with even (odd) $A$ have spin 0 ($1/2$). Only for the
deuteron the correct spin of 1 is used instead.

Further explicit
medium modifications of nuclei are not considered in HS. Since the
description of heavy nuclei is based on experimental nuclear masses, the
HS EoS includes the correct shell effects of nuclei
in vacuum. On the other hand, the use of
nuclear mass tables limits the maximum mass
and charge numbers of nuclei, see \cite{Buyukcizmeci2013}.
In such theoretical mass tables the selection of nuclei often does
not follow a consistent scheme, but is rather set by hand depending on the interest of the author. To have a
general criteria which nuclei are included, in the HS EOS calculation only nuclei
not beyond the neutron dripline are considered. As \isn{H}{4} has a negative
neutron separation energy, see Sec.~\ref{sec_4hnse}, this nucleus is \textit{not}
included in the standard versions of the HS EOS.

Later we will present results for a modified version of the HS EoS, where all possible
nuclei are taken into account, and the updated experimental measurements of \cite{Audi2014} (instead of those from \cite{Audi&Wapstra1995}) are used for the binding energies. In addition, for H and He isotopes the experimentally known excited states are explicitly included, also using the correct spin degeneracy factor (only those levels for which the spin and excitation energy are both known are considered), and the internal partition function is switched off.

\subsection{BPRS EoS}
\label{sec_BPRS}

The BPRS Equation of state, developed by \citet{Blinnikov2011},
follows most closely the conventional Saha approach \citep{Cliff&Taylor1965} and the method
by \citet{maz79}.
However,  this approach is extended in some points.
The influence of the free nucleon gas on the surface and Coulomb energies
of nuclei is taken into account.
BPRS EoS retains some terms which were omitted by \citet{maz79}, and this requires additional
loops of iteration in finding NSE.
The ``excluded-volume'' effect is neglected (the model does not pretend to reach very high densities).

There is an option to include various results for nuclear partition functions (PF)
like those of \citet{Fowler1978}
and PFs by \citet{Engelbrecht&Engelbrecht1991}.

The atomic mass table is updated using recent theoretical compilations
of atomic masses.
It covers $\sim$20000 nuclides \citep{kou07} for the KTUY mass formula \citep{kou05}
and $\sim$9000 nuclides for the FRDM mass formula \citep{Moeller1995}
as an extra option.

Some NSE models neglect the screening of the Coulomb interaction due
to the electron background, while it is accounted for in the BPRS EoS
in the Wigner-Seitz approximation.
The nuances are discussed in detail in \cite{Blinnikov2011}.

\section{\isn{H}{4} properties and NSE}
\label{sec_4hnse}
Despite the ``exotic'' status of \isn{H}{4} in calculations of the supernova context, the information about its properties is rather complete.
The binding energy per nucleon is 1.72~MeV {\citep{Audi2014}}. But earlier the accepted value was 1.394~MeV \citep{Audi&Wapstra1995}.
We'll discuss the consequences of this difference later.
Besides the groundstate there are three known excited levels of \isn{H}{4}.
Their properties (excitation energy, spin, and lifetime) are collected in  Table~\ref{tab:H4_data}.
\begin{table}
\caption{Properties of the ground state and known excited states of \isn{H}{4}, taken from \citep{Wang2014}.}
\label{tab:H4_data}
\begin{tabular}{ccc}
\hline
$\WID{E}{ex}$ (MeV) & Spin $J$ & $\tau$ ($10^{-22}$ sec)\\
\hline
0 & 2 & 1.43\\
0.31 & 1 & 0.98\\
2.08 & 0 & 0.74\\
2.83 & 1 & 0.51\\
\hline
\end{tabular}
\end{table}

It is generally accepted that unstable states must be included in the NSE approach. Here we just want to
present some additional arguments to ensure the possibility for the inclusion of \isn{H}{4} isotope into the NSE approach. The reason for this special consideration is that \isn{H}{4} has a negative neutron separation energy (although positive binding energy) and very short lifetime.

\isn{H}{4} ordinarily breaks by neutron emission with the characteristic time $\WID{\tau}{H4}\approx 1.4\times 10^{-22}$~sec  \citep{Wang2014}.
Three known exited states of \isn{H}{4} have the same order of magnitude lifetime (see table \ref{tab:H4_data}).
The simplest way to create \isn{H}{4} is the neutron capture reaction:
\begin{equation}
n+\isn{H}{3}\rightarrow \isn{H}{4} \label{reaction}
\end{equation}
But we must stress here, that this reaction (\ref{reaction}) is of course not the only way for generating \isn{H}{4}. The examples of other reactions are:
$\isn{H}{2}(\isn{H}{3},p)\isn{H}{4}$ and $\isn{H}{3}(\isn{H}{3},\isn{H}{2})\isn{H}{4}$, see~\citet{Sidorchuk2004}.
Beside this, there would be much more production and destruction channels in the hot and dense medium, e.g. fission or multifragmentation of heavy nuclei.
Furthermore, \isn{H}{4} is only abundant where there are a lot of free neutrons, so some of the decays would be blocked. Because the typical reaction times scales are very short and there are a lot of channels for creating \isn{H}{4} isotope, we conclude that its inclusion into the NSE network is well established and chemical equilibration is reached.

\section{Infall stage of collapse}
\label{sec_Infall}
\begin{figure}
	\includegraphics[width=\columnwidth]{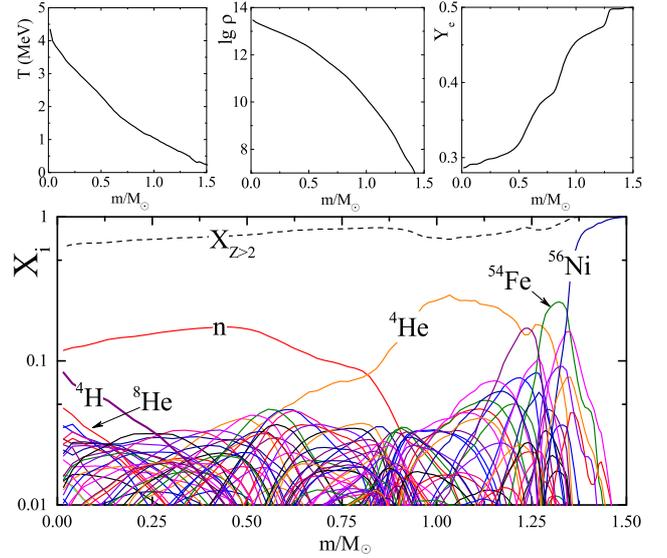}
    \caption{Upper three panels, from left ro right: temperature $T$ (in MeV), log of density
    $\rho$ (in $\mbox{g}\cdot\mbox{cm}^{-3}$) and electron fraction $\WID{Y}{e}$ as a functions of mass coordinate $m$.
    Lower panel: mass fractions of of nuclei $X_i$ as a function of $m$. The black dashed line marked
    $\WID{X}{Z>2}$ shows the total mass fraction of elements with $Z> 2$. EoS is pure NSE.}
    \label{fig:star_profile}
\end{figure}

As an example we calculate with pure NSE EoS the chemical composition of matter for the moment of time 1 ms before bounce approximately, when
the central density of a collapsing stellar core reaches roughly $3\times 10^{13}~\mbox{g}\cdot\mbox{cm}^{-3}$.
The progenitor $15~M_\odot$ stellar model is taken from \cite{Woosley&Weaver1995}. The upper three panels of Fig.~\ref{fig:star_profile} show the profiles of temperature $T$ (in MeV units),
$\log$ of density $\rho$, and electron fraction $\WID{Y}{e}$ inside the central part of the collapsing stellar core,
taken form the simulations of \citet{Hempel2012}. For the $\WID{Y}{e}$ behavior on the
pre-bounce phase see also \citet{Liebendoerfer2005}. The lower wide panel shows the profile of the chemical composition (mass fractions $X_i$)
for the same moment of time. The black dashed line marked as $\WID{X}{Z>2}$ also shows the total mass fraction of all elements with $Z> 2$.
In the outer regions with mass coordinate ($m\sim 1.5$~$M_\odot$) the core consists mainly of heavy iron-peak nuclei, while in the central, neutronized part
the \emph{individual} mass fractions of light nuclei are dominant. Besides free neutrons there are mostly helium and hydrogen isotopes.
But we need to emphasize, that the \emph{total} mass fraction of heavy nuclei is greater than 0.5 even there.
The most peculiar thing is a high abundance of \isn{H}{4} (shown by thick purple line) in the central region which is usually not taken into account.
Its mass fraction reaches almost 10\%
and it can be more abundant than alpha particles and free protons whose mass fraction is below $10^{-2}$ here. Below we analyze this \isn{H}{4}--effect in detail.

\subsection{Variation of \isn{H}{4} parameters}
Let us explore the importance of available experimental information about \isn{H}{4}, presented above in section~\ref{sec_4hnse}.
\begin{figure*}
	\includegraphics[width=2\columnwidth]{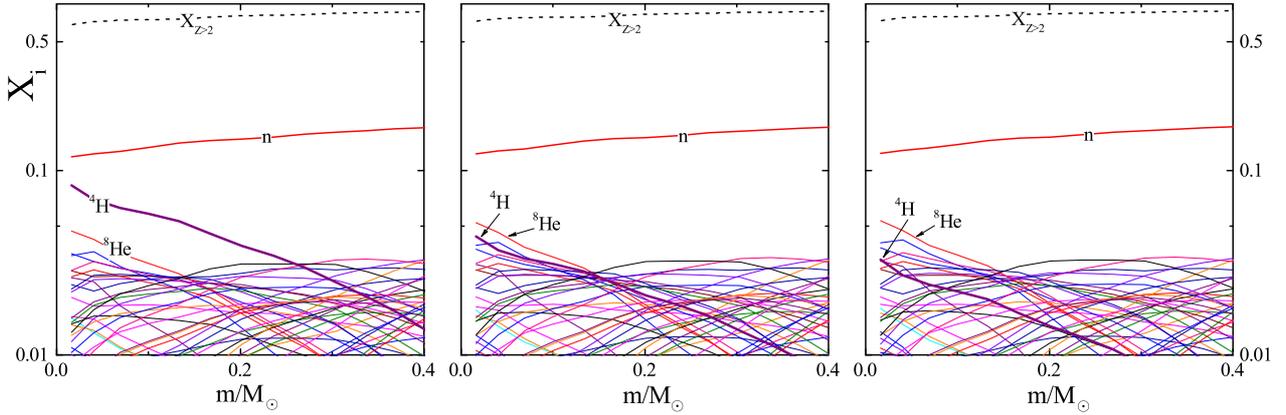}
    \caption{Mass fractions $X_i$ as a function of mass coordinate $m$.
    The star's profile, EoS and the notation are the same as in Fig.~\ref{fig:star_profile}.
    Left panel: base case. Middle: only ground state for \isn{H}{4}. Right panel: in addition
    the binding energy per nucleon of \isn{H}{4} is changed from 1.72 \citep{Audi2014} to the old value of 1.394~MeV \citep{Audi&Wapstra1995}.}
    \label{fig:H4_properties}
\end{figure*}
In Fig.~\ref{fig:H4_properties}, we plot the mass fractions of nuclei for the central part $(0\leq m/M_\odot\leq 0.4)$ of
the star's core for the same moment of time as in Fig.~\ref{fig:star_profile}. EoS again is pure NSE, please note also the restricted ordinate range.
As in Fig.~\ref{fig:star_profile}, the dashed lines marked with $\WID{X}{Z>2}$ show the total mass fraction of nuclei with $Z>2$,
$\WID{X}{H4}$ is shown  by a thick purple line.
The left panel is our base case. For the middle panel we switch off the excited levels of \isn{H}{4}, i.e., we are using only its ground state.
In the right panel in addition we use the old value of 1.394~MeV for the \isn{H}{4} binding energy per nucleon instead of 1.72~MeV.
The difference is obvious: for example, at the central point of the collapsing stellar core the mass fraction of \isn{H}{4} for the three above cases are 0.083, 0.044 and 0.033, respectively.
There are at least three factors that favor a high abundance of \isn{H}{4}, as seen in the base calculation: i) the higher value of its binding energy than previously estimated,
ii) the high value of its ground state spin of $J=2$ (see table~\ref{tab:H4_data}). In comparison with a ``naive'' prescription for the nuclear ground state
properties ($J=0$ for even $A$ and $J=1/2$ for odd $A$), \isn{H}{4} obtains a factor $2J+1=5$ benefit. And last but not least, iii) the influence of three known excited states. This illustrates the importance of using the correct experimental information for the nuclear properties. We discuss this aspect
in application to other nuclei, which is more problematic, in the next section.

\section{Influence of partition functions}
\label{sec_varpar}
At low densities the most uncertain part of the nucleus properties is its partition function (PF). Typically, one knows from the experiment only the ground state parameters and
the parameters of few (if any) low--lying exited states. To account for higher levels one has to rely on some theoretical modeling. The most widely used one is the
Fermi--gas model, which in the most simple case reduces to the Bethe--formula \citep{Bethe1936} for the nuclear PF.
More elaborated approaches consider a restricted interval of integration over the nuclear
excitation energy and a special choice of the involved parameters (see, e.g., \citet{Rauscher1997}). The most important among these parameters is the level density
parameter $a$ which is roughly proportional to the nucleus mass number $A$: $a\approx A/8~\mbox{MeV}^{-1}$. The necessary condition for using the Fermi--gas model is
$a\WID{E}{ex}\gg 1$, where $\WID{E}{ex}$ is the  total excitation energy of the nucleus. Keeping in mind that typically $\WID{E}{ex}\lesssim Q$, where $Q$ is the nucleus binding energy, we can rewrite this condition using the binding energy per nucleon
$q$, $q=Q/A$, as $A^2 q/8\gg 1$. Because for most nuclei $q\lesssim 8$~MeV, we obtain the condition for using the Fermi-gas model for the nuclear PF in the form $A\gg 1$. This results in our way how to handle partition functions: for light nuclei we use all the available experimental
information about excited states only, where we take the data from the online database of the National Nuclear Data Center, \url{http://www.nndc.bnl.gov/}.  For example, the \isn{He}{4}
PF includes all 15 known levels. For heavy nuclei (defined by the condition $Z\geq 6$, i.e., starting from carbon)
the sum over known low--lying levels is supplemented by an integral over a Fermi--gas level density. Like in HS EoS, we integrate it only up to the nucleus binding energy $Q$ to avoid the inconsistencies, see e.g. the discussion in~\cite{Yud&Nad2004}.

\begin{figure}
	\includegraphics[width=\columnwidth]{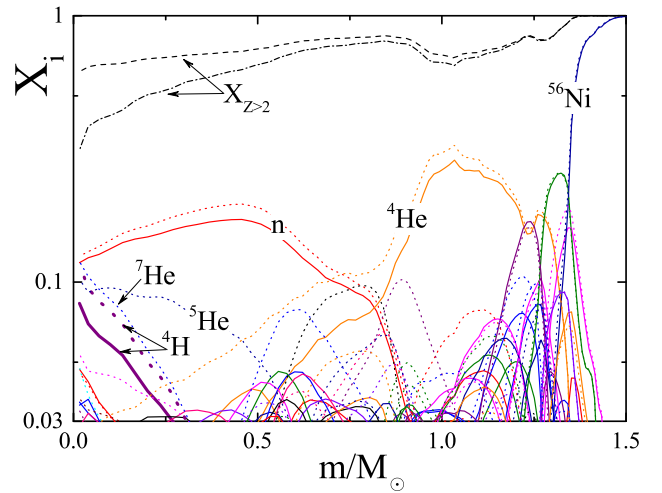}
    \caption{Mass fractions $X_i$ as a function of mass coordinate $m$.
    The star profile is the same as in Fig.~\ref{fig:star_profile}.
    Solid lines are our base case, dotted lines represent the calculation with only the
    ground state included. Dashed and dashed-dotted lines are the same for the sum of nuclei with $Z>2$. EoS is pure NSE.}
    \label{fig:compare_pf}
\end{figure}
Now we can explore the effect of the nuclear PF on the chemical composition of matter and especially on the abundance of \isn{H}{4}.
In Fig.~\ref{fig:compare_pf} we plot mass fractions $X_i$ as a function of the mass coordinate $m$ for the same star profile
 as in Fig.~\ref{fig:star_profile} (please note reduced ordinate range). Solid lines are our base case (with full PF), dotted lines show calculations with only the ground state included. As before, the dashed black line shows the total mass fraction of heavy nuclei for the base case, and the dashed-dotted line is the same for the case without excited states. As one can see the effect is quite strong. Most prominent is the radical reduction of
 the abundance of heavy nuclei at high temperature and densities. This is a well--known effect of the PF: the account for excited states of heavy nuclei permits these heavy nuclei to survive in supernova matter up to nuclear densities (see e.g. \cite{Mazurek1980}) . We see also some redistribution in
 the light nuclei abundances at the center of the collapsing stellar core: with only ground states included, the individual mass fraction of \isn{He}{7} and \isn{He}{5} dominates.
 But the most important thing for us is that \isn{H}{4} is still among the most abundant light nuclei.

 We can also consider the opposite limiting case: when the integral over excitation energy in the PF goes up to infinity.
 This is, no doubt, an overestimation of the contribution of nuclear exited states. For example, in this case the
 average excitation energy of a nucleus $\langle\WID{E}{ex}\rangle$ can significantly exceed its binding energy at high temperatures.
This is because of the rapid growth of the excitation energy with temperature: $\langle\WID{E}{ex}\rangle\approx aT^2$ (see \citet{Yud&Nad2004}).
Differences due to a cut-off for the maximal excitation energy become apparent only at high enough temperatures $T\gtrsim 10$~MeV. For the conditions
 considered here, we found negligible differences in comparison with our base case.

\section{Various EoS comparison}
\label{sec_varEOSs}
To ensure the importance of light clusters and especially \isn{H}{4} for supernova matter, we need to check this effect with other EoSs.
Let's start from the collapse situation familiar to us from Fig.~\ref{fig:star_profile}.
\begin{figure*}
	\includegraphics[width=2\columnwidth]{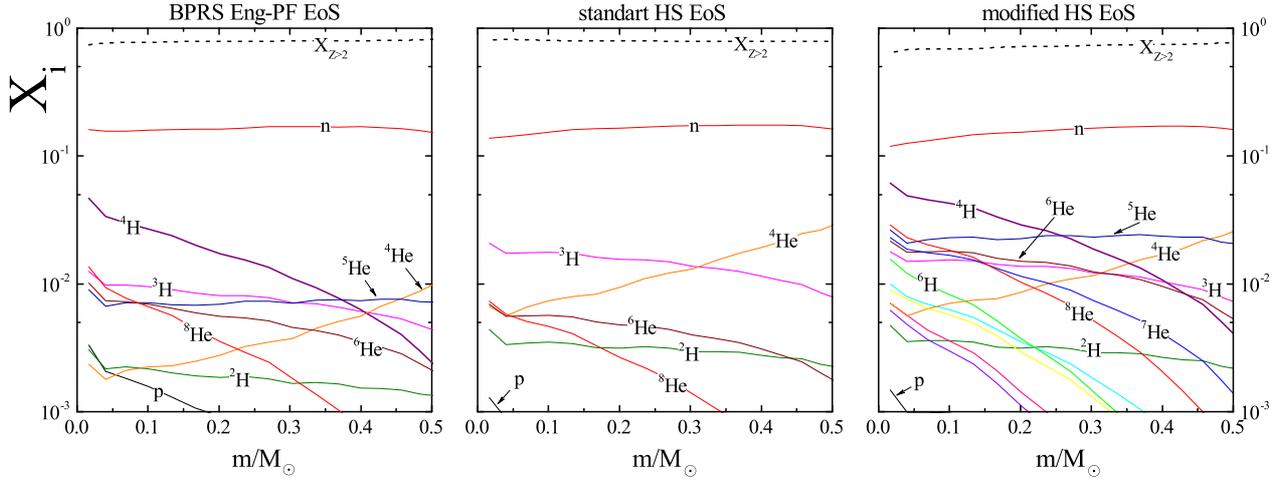}
    \caption{Mass fractions $X_i$ of \isn{H}{} and \isn{He}{} isotopes as a function of mass coordinate $m$.
    Black dashed lines also show the total mass fraction of all nuclei with $Z>2$.
    The star's profile is the same as in Fig.~\ref{fig:star_profile}.
    Left panel: BPRS EoS. Middle: standard HS EoS. Right panel: modified HS EoS.}
    \label{fig:BPRS-HS-star}
\end{figure*}
 In Fig.~\ref{fig:BPRS-HS-star} we plot the mass fractions $X_i$ of \isn{H}{} and \isn{He}{} isotopes as a function of the mass coordinate $m$.
Black dashed lines also show the total mass fraction of all nuclei with $Z>2$.
The star's profile is the same as in Fig.~\ref{fig:star_profile}. For the left panel we implemented BPRS EoS with Engelbrecht's partition functions
(BPRS Eng--PF, see subsection~\ref{sec_BPRS} for description).
For the middle we used the standard HS EoS with the restricted nuclei set (i.e. without \isn{H}{4} etc) and with the simple prescription
for nuclear levels parameters, as described above in subsection~\ref{sec_HS}. The right panel shows the results calculated with the modified HS EoS:
here we add asymmetric light isotopes and correct the spin information. It is clear that the \isn{H}{4}--effect persists even for much more elaborated
EoSs than just simple ideal NSE we used earlier.

\begin{figure*}
	\includegraphics[width=2\columnwidth]{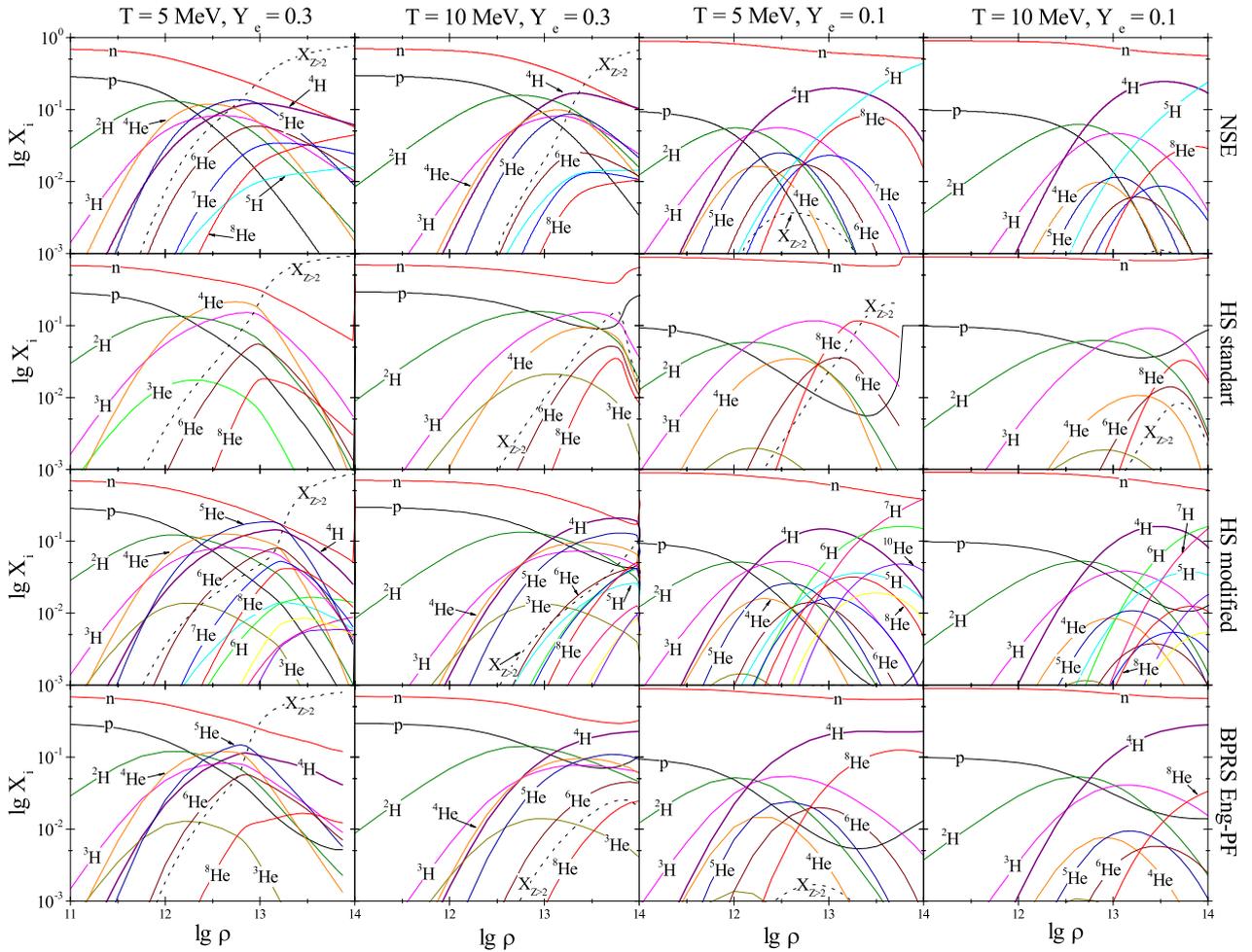}
    \caption{Mass fractions $X_i$ of \isn{H}{} and \isn{He}{} isotopes and the total mass fraction of all nuclei
    with $Z>2$ (black dashed) as a function of density. Each column corresponds to fixed temperature $T$ and lepton fraction $\WID{Y}{e}$
    shown above. The rows correspond to (from top to bottom): pure NSE, HS--standard, HS--modified and BPRS Eng--PF EoSs.}
    \label{fig:Total-EoS-Compare}
\end{figure*}

To make sure that the considered effect of light asymmetric clusters has a general application, we decided to perform an EoS comparison in a wide
thermodynamic parameter range, specific for supernova matter. The results are drawn on the composite graph~\ref{fig:Total-EoS-Compare}.
Here we plot mass fractions $X_i$ of \isn{H}{} and \isn{He}{} isotopes and the total mass fraction of all nuclei
with $Z>2$ (black dashed) as a function of density. Each column corresponds to fixed temperature $T$ and lepton fraction $\WID{Y}{e}$
shown above. We consider two temperature values $T=5$ or $10$~MeV and two $\WID{Y}{e}$ values: $0.3$ or $0.1$. As we know, $T=5$~MeV and
$\WID{Y}{e}=0.3$ conditions are characteristic for the central part of a collapsing stellar core around the bounce time.
Hotter and more neutronized conditions $T=10$~MeV and $\WID{Y}{e}=0.1$ are well suited for describing the long--time post--bounce evolution of matter
(see also section~\ref{sec_post_bounce} below). The rows correspond to (from top to bottom): pure NSE, HS--standard, HS--modified and BPRS Eng--PF EoSs, respectively. The most prominent trend of the figures, corresponding to different EoSs is their order-of-magnitude coherence. The obvious exception is HS--standard
EoS in most of the cases. For example, comparing two versions of HS EoS for $T=5$~MeV, $\WID{Y}{e}=0.1$ we see that even the
density of phase transition to nuclear matter is changed: for HS--std it occurs below $10^{14}~\mbox{g}\cdot\mbox{cm}^{-3}$ (see the sharp growth of the proton concentration)
but for HS--mod still there are a lot of nuclei even at $\rho=10^{14}~\mbox{g}\cdot\mbox{cm}^{-3}$. We need to emphasize here, that only HS-EoS is applicable at the entire density range: contrary to pure-NSE and BPRS EoSs it incorporates an excluded-volume mechanism (see, e.g. \cite{Hempel2011}) that forces nuclei to dissolve and ensures the phase transition to uniform nuclear matter at high density values. Pure NSE is good up to approximately
$\rho=10^{13}~\mbox{g}\cdot\mbox{cm}^{-3}$; at higher densities it predicts, as a rule, too low proton concentration. Also visible is the too high abundance of $X_{Z>2}$ for the
$T=10$~MeV, $\WID{Y}{e}=0.1$ case. The predictions of BPRS and HS--mod EoSs are in a good qualitative agreement for all the cases under consideration. Again we need to remind that
our main purpose here is not to perform a careful comparison of the EoSs which are quite different in the underlying physics, but rather to confirm the light clusters (and \isn{H}{4} in particular) effect.
With the above consideration we conclude that this effect is valid in a wide domain of thermodynamic
conditions for the EoSs of different nature.

Another important feature, that can be seen here on HS--mod EoS panels for the low value $\WID{Y}{e}=0.1$, are the high abundances of very neutron--rich isotopes of \isn{H}{} and \isn{He}{} (e.g. \isn{H}{7} and \isn{He}{10}) at high density. These isotopes were not included into the nuclei sets for other EoSs used. This seems to be an even more weird finding than our \isn{H}{4}--effect, but we found an independent confirmation for it. In the work \cite{Gulminelli&Raduta2015}
authors found the dominance of various exotic light nuclei such as \isn{H}{7}, \isn{He}{14}, \isn{Li}{17} etc. at high densities. Thus we believe that this effect is real and demands, no doubt, additional careful investigation.

\subsection{Partition functions for BPRS EoS}
\begin{figure}
	\includegraphics[width=\columnwidth]{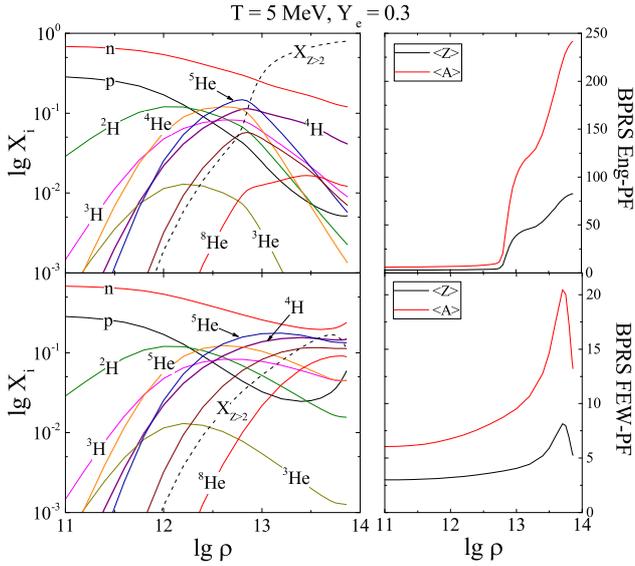}
    \caption{Mass fractions $X_i$ of \isn{H}{} and \isn{He}{} isotopes and the total mass fraction of all nuclei
    with $Z>2$ (black dashed) as a function of density. $T=5$~MeV, $\WID{Y}{e}=0.3$, EoS is BPRS. Upper case is for Eng--PF,
    lower is for FEW--PF. Right panels show corresponding average charge  $Z$ and mass number $A$ of nuclei with $Z>2$.}
    \label{fig:BPRS-pf}
\end{figure}

Before proceeding further we would like to discuss shortly the effect found during the BPRS EoS investigation for the calculations, presented in the previous section.
As described in corresponding place above (section \ref{sec_BPRS}), the BPRS EoS has two possibilities for the nuclear partition functions:
first, our base case are the PF of \cite{Engelbrecht&Engelbrecht1991} (Eng--PF). Second one are the partition functions from \cite{Fowler1978} (FEW--PF). The most prominent difference
between these approaches comes from the fact that \cite{Fowler1978} integrated nuclear level density up to the excitation energy on the order of nucleon separation energy only (their eq. (25)). This causes a serious underestimation of PF values compared to an ordinary approach. As an example we show the chemical composition of matter as a function of density for $T=5$~MeV, $\WID{Y}{e}=0.3$ for this two
approaches to PF in the Fig.~\ref{fig:BPRS-pf}.
Here, like in Fig.~\ref{fig:Total-EoS-Compare}, we plot mass fractions $X_i$ of \isn{H}{} and \isn{He}{} isotopes and the total mass fraction of all nuclei
with $Z>2$ (black dashed) as a function of density for $T=5$~MeV, $\WID{Y}{e}=0.3$. Upper case is for Eng--PF,
lower is for FEW--PF. The right panels show the corresponding average charge  $Z$ and mass number $A$ of nuclei with $Z>2$. As one can see, the difference at high densities
$\rho\gtrsim 10^{13}~\mbox{g}\cdot\mbox{cm}^{-3}$ is dramatic. High values of \cite{Engelbrecht&Engelbrecht1991} PFs permits a heavy nuclei not only to survive, but to be
a leading component of matter here. The average mass number $A$ reaches a huge value 250 approximately. Contrary to this, truncated values of PFs from \cite{Fowler1978}
lead to a suppression of heavy nuclei concentrations at high densities. Beside this, average mass number and charge are rather small here. This is the clearest example
of the importance of correct PF handling. For other combinations of temperature and lepton fraction, used in Fig.~\ref{fig:Total-EoS-Compare}, we found a much lesser effect,
probably because of the initially smaller role of heavy nuclei there. The last note concerning Fig.~\ref{fig:BPRS-pf} we should made is that the \isn{H}{4}--effect still survives
even for FEW--PF, despite of a dramatically changed chemical composition. More discussion about the role of nuclei parameters and PFs in particular, can be found in \cite{Furusawa2018}.

\section{Post-bounce calculations}
\label{sec_post_bounce}
\begin{figure}
	\includegraphics[width=\columnwidth]{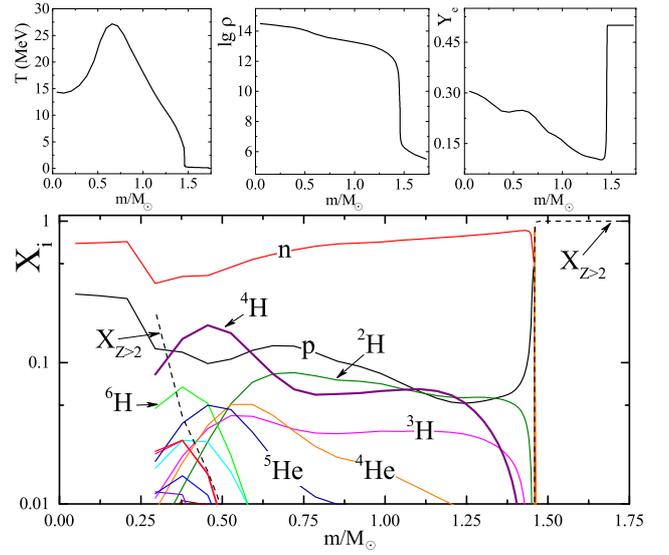}
    \caption{Upper three panels, from left ro right: temperature $T$ (in MeV), log of density
    $\rho$ (in $\mbox{g}\cdot\mbox{cm}^{-3}$) and electron fraction $\WID{Y}{e}$ as a functions of mass coordinate $m$.
    Lower panel: mass fractions $X_i$ of of hydrogen and helium isotopes as a function of $m$. The black dashed line marked
    $\WID{X}{Z>2}$ shows the total mass fraction of all rest nuclei. Stellar profile corresponds to 200 ms after bounce approximately,
    calculations according to modified HS EoS.}
    \label{fig:post-bounce}
\end{figure}
To complete our discussion about the role of light clusters and especially \isn{H}{4} for the supernova EoS, we present Fig.~\ref{fig:post-bounce}.
Here, like in Fig.~\ref{fig:star_profile} we calculate the chemical composition of matter, but for the star's profile corresponding to approximately 200~ms after bounce.
The distribution of thermodynamic parameters, shown on the upper three panels are also taken from \cite{Hempel2012}. We show only mass fractions of \isn{H}{} and \isn{He}{}
isotopes together with total mass fraction $X_{Z>2}$ of all nuclei with $Z>2$. The EoS is the modified HS, because, contrary to pure NSE and BPRS EoS it can be applied to supernuclear
densities also. The central part of the stellar core with $0\leq m/M_\odot\leq 0.25$ is occupied with pure nuclear matter made only from neutrons and protons in the strong interaction regime. The lepton fraction there is $\WID{Y}{e}\simeq 0.3$. Above this area is placed a
region passed and heated by an expanding shock wave. Behind this shock, positioned at $m\approx 1.5 M_\odot$ the matter is neutronized up to $\WID{Y}{e}\approx 0.1$.
There are a lot of light asymmetric nuclei here, in particular, of course, \isn{H}{4}. The region above the shock consists mainly of heavy nuclei with $Z>2$.
From this picture we can conclude that light clusters remain to be an important ingredient of matter during the post--bounce stage of collapse also.

\section{Importance of light clusters}
\label{sec_Importance}
\begin{figure}
	\includegraphics[width=\columnwidth]{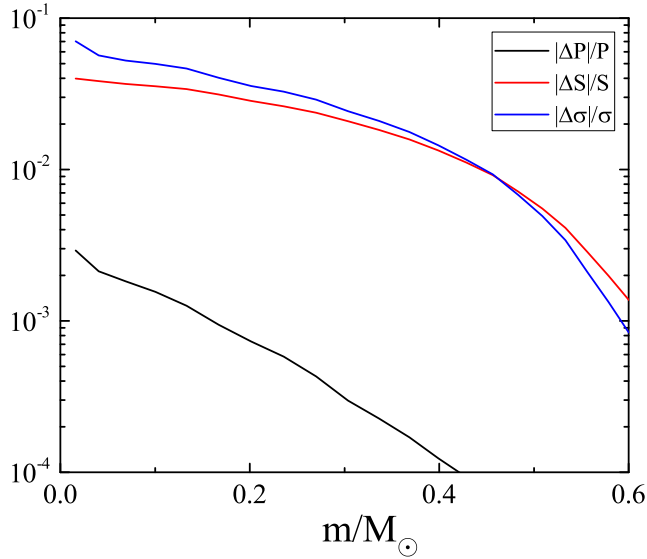}
    \caption{Relative changes of pressure (black line), entropy (red) and parameter $\sigma=\sum_i A_i^2 Y_i$.
    The underlying stellar profile is the same as in Fig.~\ref{fig:star_profile}.
    For further details see text.}
    \label{fig:deltas}
\end{figure}

After the previous discussions the main question is: after all, what is the most important effect of light clusters and in particular
\isn{H}{4} for the supernova matter problem?
Do we have to describe them accurately, or are more crude approximations sufficient?
An answer to this question can be obtained from Fig.~\ref{fig:deltas}.
Here for the same before--bounce star profile like in our Fig.~\ref{fig:star_profile} (but restricted to its central part $0\leq m/M_\odot\leq 0.6$) we plot the relative changes of a few quantities due to \isn{H}{4}.
The relative changes are
calculated with our base pure NSE EoS of Sec.~\ref{sec_NSE} with and without \isn{H}{4} included.
All other nuclei and its parameters are the same. The black line represents the relative change of the total pressure: $|\triangle P|/P$. The total pressure incudes
contributions from nuclei and free nucleons along with radiation and electron--positron pairs. One can see that the effect is negligible, or at most relatively small, well below 1\%.
The red line shows the relative change of the total entropy: $|\triangle S|/S$. In other works, see, e.g., \cite{Hempel&Schaffner2010}, it was found that the entropy is sensitive to the chemical composition. Here the effect is more pronounced than for the pressure, but still small,
only four percents in the central region of the core.

The last plotted quantity requires special explanation. An important part of the neutrino--matter interaction rates in supernovae is the coherent scattering of neutrinos on nuclei and free nucleons. Despite the fact that
this process is approximately elastic (especially for heavy nuclei), it gives an important contribution to the opacity of matter for neutrinos.
Its cross--section is roughly proportional to the \emph{square} of the nucleus mass number: $\WID{\sigma}{cs}\propto A^2$.
 Thus the total rate of coherent scattering in matter
 is proportional to $\sigma \equiv \sum_i A_i^2 Y_i=\sum_i A_i X_i$, where the sum goes
 over all nuclei and free nucleons. Here we introduce the dimensionless concentration $Y_i$ according to $Y_i\equiv \frac{n_i\WID{m}{u}}{\rho}=X_i/A_i$.
The relative change in $\sigma$ is shown by the blue line in Fig.~\ref{fig:deltas}.
 One sees that the difference in $\sigma$, caused by the presence/absence of only one nucleus \isn{H}{4} can reach 7 percent, i.e. almost twice the effect for entropy and 20 times that for pressure.
 This is because neutrino--matter interaction rates are highly sensitive to the nuclear composition.

 The main conclusions from above are the following:  thermodynamic quantities are not so sensitive to the appearance of \isn{H}{4}.
 The change in pressure and hence the direct change of the collapse dynamics will be minor. But the change of the important neutrino transport (e.g., the onset of trapping and the neutronization) and the evolution of the neutrino quantities can be important.

\section{Conclusions}
\label{sec_Conclusions}
We performed a global study of light asymmetric cluster effects for the supernova matter conditions. One of our most peculiar finding is a high abundance of the
\isn{H}{4} isotope which was not taken into account previously. We explore a wide domain of thermodynamic parameters, representative for the matter of a collapsing stellar
core during the infall stage as well as during the post--bounce phase. It appears that for light nuclei it is important to use exact information about their properties
(values of spins and energies of known excited states) to obtain a reliable EoS. For the heavy nuclei the effect of the whole partition functions occurs to be of the same importance.
By comparison of three EoSs with different underlying physics we ensured the stability of the light cluster effect for different conditions. We have found that various very asymmetric light isotopes of \isn{H}{} and \isn{He}{} can be abundant in neutron-rich matter (see e.g. the panels of Fig.~\ref{fig:Total-EoS-Compare} corresponding to $\WID{Y}{e}=0.1$ and HS-mod EoS). But the calculations with the stellar profiles from pre-bounce (Fig.~\ref{fig:star_profile}) and post-bounce (Fig.~\ref{fig:post-bounce}) phases of collapse, reveal the leading role of \isn{H}{4} nuclide among other exotic light clusters, although the traces of e.g. \isn{He}{8} or \isn{H}{6} are also visible. In these stellar conditions \isn{H}{4} can be even more abundant than the deuteron and tritium isotopes, usually included into the nuclei network.
At the end we discussed the domains in supernova modelling where the light nuclei effect can have most important consequences.

\section*{Acknowledgements}
We would like to thank Tobias Fischer for his kind permission to use the data for star profiles from his simulations.\\
Authors thank RFBR grants No 16-02-00228a and No 18-29-21019 for financial support.
Authors would like also to thank the anonymous referee for his/her numerous and very useful comments and suggestions.





\begin{thebibliography}{99}


\bibitem[\protect\citeauthoryear{Audi et al.}{2014}]{Audi2014}
Audi G., Wang M., Wapstra A. H., Kondev F. G., MacCormick M., Xu X., 2014,
Nuclear Data Sheets, 120, 1-5

\bibitem[\protect\citeauthoryear{Audi and Wapstra}{1995}]{Audi&Wapstra1995}
Audi G., Wapstra A.H. and Thibault C., 2003, Nuclear Physics A, \textbf{729}, 337

\bibitem[\protect\citeauthoryear{Bethe}{1936}]{Bethe1936}
Bethe A.H., 1936, Phys. Rev. \textbf{50}, 332

\bibitem[\protect\citeauthoryear{Blinnikov et al.}{1996}]{Blinnikov1996}
Blinnikov S.I., Dunina--Barkovskaya N.V. and Nadyozhin D.K., 1996, Astrophys. J. Suppl. \textbf{106}, 171

\bibitem[\protect\citeauthoryear{Blinnikov et al.}{2011}]{Blinnikov2011}
Blinnikov S.I., Panov I.V., Rudzsky M.A., Sumiyoshi K., 2011, Astron. Astr., \textbf{535}, 13

\bibitem[\protect\citeauthoryear{Buyukcizmeci et al.}{2013}]{Buyukcizmeci2013}
Buyukcizmeci, N., et al., 2013, Nucl. Phys. A \textbf{907}, 13

\bibitem[\protect\citeauthoryear{Clifford and Taylor}{1965}]{Cliff&Taylor1965}
Clifford~F.E., Tayler~R.J., 1965, Mem. R. Astron. Soc. \textbf{69}, 21

\bibitem[\protect\citeauthoryear{Engelbrecht and Engelbrecht}{1991}]{Engelbrecht&Engelbrecht1991}
Engelbrecht C.A., Engelbrecht J.R., 1991, Annals of Phys. \textbf{207}, 1-37

\bibitem[\protect\citeauthoryear{F\'ai and Randrup}{1982}]{Fai&Randrup1982}
F\'ai G.,  Randrup J., 1982, Nucl. Phys. A \textbf{381}, 557

\bibitem[\protect\citeauthoryear{Fischer et al.}{2014}]{Fischer2014}
Fischer T., Hempel M., Sagert I., Suwa Y., Schaffner-Bielich J., 2014, Eur. Phys. J. A \textbf{50}, 46

\bibitem[\protect\citeauthoryear{Fischer et al.}{2016}]{Fischer2016}
Fischer T., Mart\'inez-Pinedo G., Hempel M., Huther L., R\"opke G., Typel S., Lohs A., 2016, EPJ Web Conf. \textbf{109}, 06002

\bibitem[\protect\citeauthoryear{Fowler et al.}{1978}]{Fowler1978}
Fowler W.A., Engelbrecht C.A. and Woosley S.E. 1978, Astrophys. J., \textbf{226}, 984--995

\bibitem[\protect\citeauthoryear{Furusawa et al.}{2013}]{Furusawa2013}
Furusawa S., Nagakura H., Sumiyoshi K., Yamada S.,
2013, Astrophys. J. \textbf{774}, 78

\bibitem[\protect\citeauthoryear{Furusawa}{2018}]{Furusawa2018}
Furusawa S., 2018, arXiv:1811.10198, accepted to Phys. Rev. C


\bibitem[\protect\citeauthoryear{Gulminelli and Raduta}{2015}]{Gulminelli&Raduta2015}
Gulminelli F., Raduta Ad.R., 2015, Phys. Rev. C, \textbf{92}, 5

\bibitem[\protect\citeauthoryear{Hempel and Schaffner-Bielich}{2010}]{Hempel&Schaffner2010}
Hempel M., Schaffner-Bielich J., 2010, Nuclear Physics A,  \textbf{837}, 3-4,  210-254

\bibitem[\protect\citeauthoryear{Hempel et al.}{2011}]{Hempel2011}
Hempel M., Schaffner-Bielich J., Typel S., R\"opke J., 2011, Phys. Rev. C, \textbf{84}, 5

\bibitem[\protect\citeauthoryear{Hempel et al.}{2012}]{Hempel2012}
Hempel M., Fischer T., Schaffner-Bielich J., Liebend\"orfer M., 2012, Astrophys. J. \textbf{748}, 70

\bibitem[\protect\citeauthoryear{Hempel et al.}{2015}]{Hempel2015}
Hempel M., Hagel K., Natowitz J., R\"opke G., Typel S., 2015, Phys. Rev. C \textbf{91}, 045805

\bibitem[\protect\citeauthoryear{Horowitz and Schwenk}{2006}]{Horowitz&Schwenk2006}
Horowitz C., Schwenk A., 2006, Nucl. Phys. A \textbf{776}, 55

\bibitem[\protect\citeauthoryear{Horowitz et al.}{2014}]{Horowitz2014}
Horowitz C., Brown E., Kim Y., Lynch W., Michaels R.,
et al., 2014, J. Phys. G \textbf{41}, 093001

\bibitem[\protect\citeauthoryear{Imshennik and Nadyozhin}{1965}]{Imsh&Nad1965}
Imshennik V.S. and Nadezhin D.K., 1965, Astron. Zh., \textbf{42}, 1154

\bibitem[{{Koura}(2007)}]{kou07} {Koura}, H. 2007, private communication

\bibitem[{{Koura} {et~al.}(2005){Koura}, {Tachibana}, {Uno}, \&
  {Yamada}}]{kou05}
{Koura}, H., {Tachibana}, T., {Uno}, M., \& {Yamada}, M. 2005, Progress of
  Theoretical Physics, 113, 305

\bibitem[\protect\citeauthoryear{Kowalski et al.}{2007}]{Kowalski2007}
Kowalski S., Natowitz J., Shlomo S., Wada R., Hagel K., et al., 2007, Phys.Rev. C \textbf{75}, 014601


\bibitem[\protect\citeauthoryear{Langer et al.}{1969}]{Langer1969}
Langer W.D., Rosen L.C., Cohen J.M., Cameron A.G.W., 1969, \apss, \textbf{5}, 3, 259-271

\bibitem[\protect\citeauthoryear{Lattimer and Swesty}{1991}]{Latt&Swesty1991}
Lattimer~J.M., Swesty~F.D., 1991, Nucl. Phys.~A. \textbf{535}, 331-376

\bibitem[\protect\citeauthoryear{Liebend\"{o}rfer}{2005}]{Liebendoerfer2005}
Liebend\"{o}rfer M., 2005, Astrophys. J. {\bf 633}, 1042

\bibitem[\protect\citeauthoryear{Mallik et al.}{2008}]{Mallik2008}
Mallik S., De J., Samaddar S., Sarkar S., 2008, Phys. Rev.
C \textbf{77}, 032201.

\bibitem[{Mazurek {et~al.}(1979)Mazurek, Lattimer, \& Brown}]{maz79}
Mazurek, T.~J., Lattimer, J.~M., \& Brown, G.~E. 1979, Astrophys.\ J., 229, 713

\bibitem[\protect\citeauthoryear{Mazurek and Brown}{1980}]{Mazurek1980}
Mazurek T.J., Brown G.E., 1980, Astron. Astrophys. \textbf{81}, 382-386

\bibitem[\protect\citeauthoryear{M{\"o}ller et~al.}{1995}]{Moeller1995}
M{\"o}ller P., Nix J.~R., Myers W.~D., Swiatecki W.~J., 1995, Atom. Data Nucl. Data \textbf{59}, 185

\bibitem[\protect\citeauthoryear{Natowitz et al.}{2010}]{Natowitz2010}
Natowitz J., R\"opke G., Typel S., Blaschke D., Bonasera A.,
et al., 2010, Phys. Rev. Lett. \textbf{104}, 202501

\bibitem[\protect\citeauthoryear{Oertel et al.}{2016}]{Oertel2016}
Oertel M., Hempel M., Kl\"ahn T., Typel S., 2017, Rev. Mod. Phys., \textbf{89}, 1

\bibitem[\protect\citeauthoryear{O`Connor et al.}{2007}]{Oconnor2007}
OConnor E., Gazit D., Horowitz C., Schwenk A., Barnea N., 2007, Phys.Rev. C \textbf{75}, 055803

\bibitem[\protect\citeauthoryear{Pais et al.}{2015}]{Pais2015}
Pais H., Chiacchiera S., Provid\^encia C., 2015, Phys. Rev.
C \textbf{91}, 055801

\bibitem[\protect\citeauthoryear{Rauscher et al.}{1997}]{Rauscher1997}
Rauscher~T., Thielemann~F.-K., Kratz~K.L. 1997, Phys. Rev.~C., \textbf{56}, 163

\bibitem[\protect\citeauthoryear{R\"opke}{2015}]{Roepke2015}
G. R\"opke, 2015, Phys. Rev. C, \textbf{92}, 054001

\bibitem[\protect\citeauthoryear{Shen et al.}{1998a}]{Shen1998a}
Shen, H., Toki, H., Oyamatsu, K., Sumiyoshi, K. 1998, Prog. Theor. Phys.,
\textbf{100}, 1013

\bibitem[\protect\citeauthoryear{Shen et al.}{1998b}]{Shen1998b}
Shen, H., Toki, H., Oyamatsu, K., Sumiyoshi, K. 1998, Nucl. Phys. A, \textbf{637},
435


\bibitem[\protect\citeauthoryear{Sidorchuk et~al.}{2004}]{Sidorchuk2004}
Sidorchuk S.~I. et al., 2004, Phys. Lett. B,
594, 54-60

\bibitem[\protect\citeauthoryear{Sumiyoshi and R\"opke}{2008}]{Sumiyoshi&Roepke2008}
Sumiyoshi K., R\"opke G., 2008, Phys. Rev. C \textbf{77}, 055804

\bibitem[\protect\citeauthoryear{Typel et al.}{2010}]{Typel2010}
Typel, S., R\"opke G., Kl\"ahn T., Blaschke D., Wolter H., 2010,
Phys. Rev. C \textbf{81}, 015803

\bibitem[\protect\citeauthoryear{Wada et al.}{2012}]{Wada2012}
Wada R., Hagel K., Qin L., Natowitz J., Ma Y., et al., 2012,
Phys. Rev. C \textbf{85}, 064618

\bibitem[\protect\citeauthoryear{Wang et~al.}{2014}]{Wang2014}
 Wang, M., Audi  G., Kondev  F. G., Pfeiffer B., Blachot J., Sun X., MacCormick M., 2014,
Nuclear Data Sheets, 120, 6-7

\bibitem[\protect\citeauthoryear{Wang et~al.}{2012}]{Wang2012}
Wang M., Audi G., Wapstra A.~H., Kondev F.~G., MacCormick M., Xu X., Pfeiffer B. 2012, Chinese
Physics C, 36, 12, 1603-2014

\bibitem[\protect\citeauthoryear{Woosley and Weaver}{1995}]{Woosley&Weaver1995}
Woosley, S.E., and Weaver, T.A., 1995, ApJS, 101, 181


\bibitem[\protect\citeauthoryear{Yudin and Nadyozhin}{2004}]{Yud&Nad2004}
Yudin A.V., Nadyozhin D.K., 2004, Astron. Lett., \textbf{30}, 9, 634-646

\end{thebibliography}




\bsp	
\label{lastpage}
\end{document}